# Modulation of magnetism via electric field in MgO nanoribbons

Liang Liu, Xue Ren, Jihao Xie, Bin Cheng, Jifan Hu[*]

School of Physics, State Key Laboratory for Crystal Materials, Shandong University, Jinan 250100, China

[*] Corresponding author, email: hujf@sdu.edu.cn or hu-jf@vip.163.com

**Abstract**

We report on a theoretical study of electromagnetic (ME) properties of zigzag magnesium oxides nanoribbons (Z-MgONRs). We propose that the polar charges and the spin polarization are the two key factors for edge magnetism. Based on first-principle calculations, we demonstrate that both O- and Mg-edges are magnetic and their magnetic moments all can be efficiently modulated via external electric field. And the edge magnetism is further studied in the framework of effective tight-binding model, which provides a starting point for the calculation of one particle Green's function and the determination of exchange interactions. Utilizing the linear response model, we find that Z-MgONRs exhibit two kinds of exchange interaction with extremely different natures. The magnetism in O-edge is strongly localized with ferromagnetic order while the magnetism in Mg-edge is itinerant with Ruderman-Kittel-Kasuya-Yosida (RKKY) like interactions. And these two couplings can also be modulated by electric field, giving rise to the electrical modulations of spin-density-wave (SWD) and the modulation of Curie temperature in O-edge. All these suggest that Z-MgONRs are ideal platforms for ME coupling and appealing candidates for manifold applications.

## Introduction

The realization of magnetism in two-dimensional (2D) systems is amongst the most topical issues in nowadays. Due to the high boundary-bulk ratio in 2D system, the intrinsic magnetism may be dominated by the boundary states which are sensitive to external stimulus[1-9]. Such properties pave a new way to efficiently control the magnetism, which is highly desired for the ever-growing technical demanding[10-13]. The discovering of room temperature ferromagnetic zigzag graphene nanoribbons (ZGNRs) suggests one possible way to induce 2D magnetism through edge effects[14]. By doping, edge-decoration and applying external electric field, the magnetism in ZGNRs can be efficiently modulated and even half metallic states can be realized[15-21]. To date, the edge engineering has become one of the most promising ways to induce and modulate magnetism in 2D systems such as $MoS_2$, hexagonal BN, ZnO and so on[22-27].

MgO is a famous simple oxide with rock salt structure in bulk phase. The 2D MgO with both polar (111) and nonpolar (001) orientations have been successfully growth on various substrates in experiments[28-34]. Particularly, the polar orientation i.e. MgO(111) present sharp interests due to its peculiarities and exotic electronic, magnetic and optical properties[35-43]. Both theoretical and experimental researches suggest that the 2D MgO(111) possesses a graphene-like planar honeycomb structure, thus eliminates the out-plane polarity and structural instability. Similar to other 2D systems with binary components, the non-centrosymmetric structure and the ionic nature lead to an intrinsic bulk polarization **P** along the Mg-O bond direction. This nontrivial polar effect is a remarkable property that enables polar charges in zigzag edges, forming quasi one-dimensional electron gas (1DEG) and the high density of edge states at Fermi level. This further gives rise to the exchange splitting according to Stoner criteria[38]. Therefore, it is promising to induce magnetism into 2D MgO(111) via zigzag edge effects.

Nevertheless, the study on the magnetic structure of Z-MgONRs is still on its infancy. Former researches are focused on the existence and stability of edge magnetism, the mechanism of exchange interactions has not been well established. In addition, no research has surveyed how the edge magnetism responds to external stimulus especially external electric field, which is essential for a wide range of future applications. Hence, further effort is required to obtain a deeper understanding and better controlling of the edge magnetism in Z-MgONRs.

In this work, we report a comprehensive study on the edge magnetism in Z-MgONRs. There are two primary aims of this study: to investigate the ME effects and to ascertain the magnetic coupling mechanism. Firstly, we propose that the polar charges driven by the nontrivial bulk polarization **P** and the spin polarization edge states are the two key factors for ground-state magnetism **M**. Based on extensive density functional theory (DFT) calculations, we show that both O- and Mg-edge are magnetic but with different natures. O-edge is thoroughly spin polarized and the **M** is simply proportional to external transverse electric field **E**. However, Mg-edge is partially spin-polarized and the **M** is reduced by applied **E**. Secondly, by representing the electronic states in terms of tractable tight-binding model and utilizing single-particle Green's function, we find that the magnetism in O-edge has a ferromagnetic order with short range exchange interactions, while the exchange interactions in Mg-edge are long range and show oscillatory characters which are similar to the RKKY mechanism. It is a significant result that the Z-MgONRs simultaneously support two kinds of exchange mechanism with extremely different natures. Furthermore, we demonstrate that the exchange interactions are also amenable to external **E**, giving rise to the electrical modulations in SDW spectra. Furthermore, we perform a series of Monte Carlo simulations and verify that the Curie temperature in O-edge can be sufficiently modulated via **E**, too. These findings enhance our knowledge of edge-magnetism in Z-MgONRs and shed light on the realization of ME couplings and will expand the future application of MgO.

## Methods

We performed DFT calculations using VASP code package[44, 45]. The exchange and correlation of electrons are described utilizing Perdew-Burke-Ernzerhof form of generalized gradient approximation, and the projected-augmented-wave pseudo-potential is used for the description of core and valence electrons. The kinetic energy of plane-wave expansion is < 400 eV. A 12×1×1 k-points grid using Monkhorst-Pack method is used for sampling Brillouin zone. And The energy tolerance for the self-consistent calculations is set to $10^{-5}$ eV. Both the atom positions and lattice constant are allowed to relax until the maximal force on each atom <$10^{-2}$ eV/Å. At least 15 Å vacuum spaces were involved in each non-periodic direction for the elimination of ribbon-ribbon interactions. Wolff algorithm is used in Monte Carlo simulations, showing excellent efficiency in the near-transition region. And we use 1000 unit cells with periodic condition for the simulation. And more than $10^6$ sweeps are used for statistics [46].

## Results and discussion

### 1. Modulations of magnetic moment

The properties of Z-MgONRs are deeply inherited from their parent system: 2D MgO(111) who owns a hexagonal lattice composing of two distinct 2D sublattices of Mg and O atoms respectively. These two unequal sub-lattices leads to an overall point group of $C_{3v}$ and leads to the bulk polarization[47, 48]. The computation of this polarization is straightforward once we represent the electronic structure with the basis of maximally localized Wannier functions (MLWF)[49]:

$$\boldsymbol{P} = \frac{e}{A}\left(\sum_i Z_i \boldsymbol{l}_i - \sum_j \langle \boldsymbol{r} \rangle_j\right) + \frac{e}{A}\boldsymbol{R} \qquad (1)$$

here, $Z_i$ is the valence state of the ith ion, $\boldsymbol{l}_i$ is the position of the nucleus, $\langle \boldsymbol{r} \rangle_j$ represents the charge center of MLWFs, and $A$ is unit cell (u.c.) area, and $\boldsymbol{R}$ can be any Bravais-lattice vector since the formal polarization $\boldsymbol{P}$ is a lattice of vectors rather than a single vector, according to the modern theory of polarizations[50-53]. In the case of MgO(111), due to the particularly ionic nature, the contribution of oxygen ion core can be simply incorporated with eight valence, forming a Wannier anion[53]. And we can only consider the contribution of magnesium ion and Wannier anion, since the $C_{3v}$ symmetry eliminates the dipole of Wannier anion. Hence, the charge number of our Wannier anion is -2 and $\boldsymbol{l}_W = \boldsymbol{l}_O$, and the total polarization of appropriate bulk cell can be easily calculated according to eq. 1:

$$\boldsymbol{P} = -\frac{e}{3A}(\boldsymbol{a_1} + \boldsymbol{a_2}) = -\frac{2e}{3a}\hat{n} \qquad (2)$$

here $\hat{n}$ is a unit vector parallel to the armchair direction and we have set $\bm{R} = 0$ as a specific representative element of polarization defined in eq. 1. Therefore, the polarization $\bm{P}$ points the armchair direction, in agreement with the imposed $C_{3v}$ symmetry restriction[54, 55].

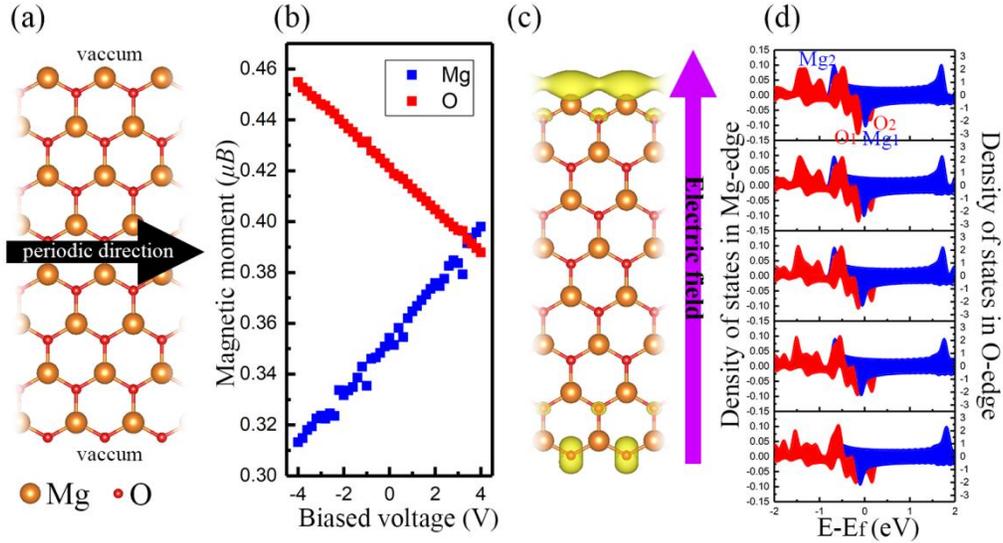

**Figure 1**. (a) the relaxed geometry structure of Z-MgONRs. (b) the biased voltage dependent magnetic moments. The red and blue squares represent the magnetic moment of O- and Mg-edges, respectively. (c) the free-field spin-density distribution in ground state, the isosurface of spin-density is set to 0.006 Å$^{-3}$. (d) DOS diagram under biased voltage ΔV. From top to bottom, ΔV is equal to 4V, 2V, 0V, -2V and -4V, respectively. The arrow at left side defines the positive direction of applied external electric field.

Z-MgONRs only preserve the zigzag periodic direction of MgO(111), the relaxed geometrical structure has been shown in Fig. 1(a). Since the bulk polarization **P** in MgO(111) owns the same orientation of the symmetry-breaking, the build-in electric field is induced and the edge states have been pushed to the near of Fermi level. Due to the strong locality of these edge states, they are presented as high peaks in the density of states (DOS) diagram i.e. the von Hove singularities. Fig. 1(d) shows four such peaks including two for Mg-edge and two for O-edge. On the other hand, polar charges are induced to screen the intrinsic electric field, avoiding the polar catastrophe[53, 56]:

$$\frac{\sigma}{l} = \hat{n} \cdot \Delta \boldsymbol{P} \qquad (3)$$

where $\sigma$ denotes the quantity of polar charges and $l$ denotes the length of boundary, in Z-MgONRs case, $l$ is simply equal to the lattice constant a, $\hat{n}$ is the unit vector orthogonal to the edge pointing to the vacuum, $\Delta \boldsymbol{P}$ is the difference of polarization across the edge. The polarization of vacuum can be treated as zero, thus $\Delta \boldsymbol{P}$ is identical to $\boldsymbol{P}$ defined in eq. 2, and we obtain the polar charges arised in the edges: $|\sigma_{zz}| = \frac{2e}{3}$. The negative sign indicates the electron-doping in Mg-edge and the positive sign indicates hole-doping in O-edge. In the view of DOS filling, this kind of charge doping is expressed by the hole occupation in valence bands of O-edge, corresponding to $O_2$ peak, and the electron occupation in conduction bands of Mg-edge, corresponding to $Mg_1$ and $Mg_2$ peak. The van Hove singularities with partially filling induce the Stoner instability and give rise to the edge magnetism. Now we can associate the polar charges $\sigma_{zz}$ with the spin-polarized DOS peaks:

$$|\sigma_{zz}| = \int_{-2}^{0}(-O_1 - O_2)dE = \int_{0}^{2}(-Mg_1 + Mg_2)dE \qquad (4)$$

we only consider the energy range from -2 eV to 2 eV because all meaningful edge states in this study are located in this range. Since the magnetic moments of edge O ($\mathbf{M_O}$) and Mg ($\mathbf{M_{Mg}}$) ions are proportional to the occupation of $O_1 + O_2$ and $Mg_1 + Mg_2$, respectively, we naturally expect that $\mathbf{M_O}$ should be equal to $|\sigma_{zz}|$ and $\mathbf{M_{Mg}}$ should be smaller. However, it must be noted that the correctness of eq. 3 is based on the hypothesis that the polar charges would thoroughly eliminate the intrinsic electric field. In fact, polar catastrophe only emerges in infinite large systems. Our Z-MgONRs with finite size always sustains a bit of residual polarizations. Hence, $|\sigma_{zz}|$ is always smaller than the maximum value of $\frac{2e}{3}$. In addition, not all of the polar charges are strictly localized in the outer most ions, too. Given all these above, even in the O-edge with complete spin-polarization edge states, the local magnetic moment is still below $\frac{2}{3}\mu B$, as shown in Fig. 1(b).

A transverse electric field will give rise to extra polar charges due to the dielectric effect:

$$\frac{\sigma}{l} = \hat{n} \cdot \Delta(\boldsymbol{P} + \varepsilon \boldsymbol{E}) \Longrightarrow \Delta\sigma_{zz} = a\varepsilon\hat{n} \cdot \boldsymbol{E} \qquad (5)$$

here, $\varepsilon$ is the permittivity of Z-MgONRs and we assume that it does not rely on **E**. As shown in Fig. 1(c), the defined positive direction of **E** is opposite to **P**. Therefore, the polar charges are drained linearly with the magnitude of electric field. Since all the polar charges in O-edge contribute to the magnetism, the magnetic moment of O-edge is also reducing linearly as the electric field increases. In terms of DOS evolution which is given in Fig. 1(d), the applied electric field generates an extra electrostatic potential to edge states, and lowers the energy level of O-edge, thus the hole-filling in $O_2$ peak is reduced and the magnetic moment is weakened.

On the other hand, the evolution of magnetic moment of Mg-edge is thoroughly different, as one can see in Fig. 1(d). In this case, albeit the polar charges are still reduced by **E**, the spin-polarization is enhanced. The competition of these two factors leads to a more complex behavior. The DOS diagram shown in Fig. 1(d) reveals that the Fermi level crosses the $Mg_1$ peak which negatively contributes to the magnetic moment. The electron-filling of $Mg_1$ peak decreases rapidly as the extra electrostatic potential pulls the state of Mg-edge to the higher energy region. Hence, crudely speaking, the magnetic moment of Mg-edge is enhanced by **E**.

## 2. Magnetic coupling mechanism

Within the formalism of DFT based on periodic boundary conditions, we can hardly determine the ground-state magnetic structure specially the long range magnetic order, since such a long range property needs to be computed using very large supercell, which is too expensive for DFT calculations. Alternatively, we study the magnetic properties on the basis of classic Heisenberg model:

$$H = -\sum_{i,j} J_{ij} \mathbf{e}_i \mathbf{e}_j \qquad (6)$$

where $J_{ij}$ denotes the exchange interactions between i[th] and j[th] sites, and $\mathbf{e}_i$ is the unit vector in the direction of the magnetic moment of i[th] site. Positive $J_{ij}$ indicates the ferromagnetic (FM) order between i[th] and j[th] sites is energetically favored, while negative $J_{ij}$ reveals the anti-ferromagnetic (AFM) order. It is also possible that the exchange interactions in some systems involve both positive and negative parts, e.g. the RKKY coupled systems. In order to use this approach, we first represent the ground states in the framework of tight-binding (TB) model and obtain the effective Hamiltonian. Then we can calculate the one-particle Greens' function and determine the exchange parameters in Heisenberg model.

Now let us trace the electronic structures in terms of TB model. The Hamiltonian is:

$$H_{\uparrow(\downarrow)} = \sum_i \varepsilon_i^{\uparrow(\downarrow)} c_{i\uparrow(\downarrow)}^\dagger c_{i\uparrow(\downarrow)} + \sum_{i,j} t_{ij}^{\uparrow(\downarrow)} c_{i\uparrow(\downarrow)}^\dagger c_{j\uparrow(\downarrow)} \qquad (7)$$

here, the arrows represent the spin-up or down tunnel, $\varepsilon_i$ is the onsite energy of i[th] site. For each given site, the energy difference between spin up and down tunnel is the exchange energy: $\Delta_i = \varepsilon_i^\uparrow - \varepsilon_i^\downarrow$. It was found that only the orbitals of edge O and Mg ions have nonzero exchange energies, in agreement with former discussions. $c_i^\dagger(c_i)$ is the field operator which creates (annihilates) one electron at i[th] site, respectively. And $t_{ij}$ is the amplitude of hopping between i[th] and j[th] sites. In order to sufficiently reproduce the electronic states of Z-MgONRs, the TB basis involves a large contribution of orbitals of edge Mg ions. These orbitals have relatively nonlocal features, spreading over the Mg-edge, we rather consider the cutoff in $t_{ij}$ i.e. ignore all hopping terms which fulfill the condition $|t_{ij}| < 0.02$ eV. All parameters in eq. 6 can be extract from the ground states calculated by DFT[49].

The most straightforward way to verify the efficiency of our TB Hamiltonian is to check the dispersion law. Utilizing standard lattice Fourier transformations to the real-space Hamiltonian defined in eq. 1, we can get the reciprocal Hamiltonian $H_k^{\uparrow(\downarrow)}$ and solve its eigenvalues. In Fig. 2, we compare the band spectra calculated by DFT and our TB model, and one can see the effective TB model excellently describes the electronic states for the cases with or without biased-voltage, at least in the energy range of $\pm 2$ eV around Fermi level which domains the magnetic properties. Fig. 2 also tells us the story of electrical modulation in the version of band structures. Biased voltage pushes the spin-polarized bands of edge Mg and O especially $O_1$ and $Mg_2$ bands to lower or higher energy, inducing the variation of spin-polarized charges and magnetic moment.

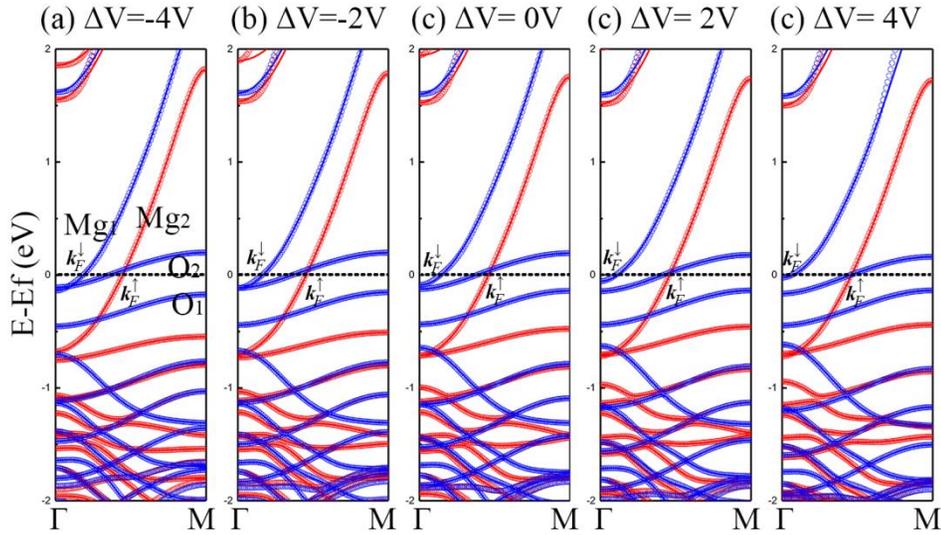

**Figure 2**. Spin-resolved band structures under each biased voltage. Red and blue colors correspond to the spin-up and spin-down tunnels, respectively. Thick lines are calculated via DFT while open circles are calculated via effective TB model. Dashed black lines denote the Fermi level. $k_F^\uparrow$ and $k_F^\downarrow$ denote the Fermi wavevectors of $Mg_1$ and $Mg_2$ bands, respectively.

With the expression of reciprocal $H_k^{\uparrow(\downarrow)}$, we can compute the retarded single-particle Green's function for each spin tunnel:

$$G_k^{\uparrow(\downarrow)}(z) = \frac{1}{z - H_k^{\uparrow(\downarrow)} + i\eta} \qquad (8)$$

where the constant $\eta$ is set to $10^{-4}$ eV. And via Fourier transformations we obtain the real-space expression of Green's function:

$$G_{ij}^{\uparrow(\downarrow)}(z) = \sum_{k} e^{-ik(R_i - R_j)} G_k^{\uparrow(\downarrow)}(z) \quad (9)$$

In the calculations for real-space Green's function, we use an adaptive k-mesh sampling algorithm, for the energy $z$ near to eigenvalues of $H_k$, at least $10^6$ k-points are used to get a converged result (with $|\Delta G_{ij}(z)| < 0.01\ meV^{-1}$). Then we can compute the exchange parameters $J_{ij}$ according to the linear response theorem, also called magnetic force theorem[57-59]:

$$J_{ij} = Tr^L\ Im\ \frac{1}{4\pi} \int_{-\infty}^{Ef} dz\ \Delta_i \Delta_j G_{ij}^{\uparrow} G_{ij}^{\downarrow} \quad (10)$$

here, $Ef$ is the Fermi level, $Tr^L$ denotes the trace over angular moment on $i^{th}$ and $j^{th}$ sites. In practice, we push the energy into the upper half complex zone, and set the integration route along the semi-circle contour starting at the bottom of bands and ending at the Fermi energy. The integration is solved by Legendre-Gauss quadrature integral approximation with twenty sampling points on the contour. Due to the translational periodicity of our system, the exchange parameters are only relevant to the distance between two sites, i.e. $d = |i - j|$. Therefore, we only need to consider $J_{0d}$ which is related to the "home cell". And the exchange interactions now become a discrete function of $d$, writing as $J(d)$ concisely.

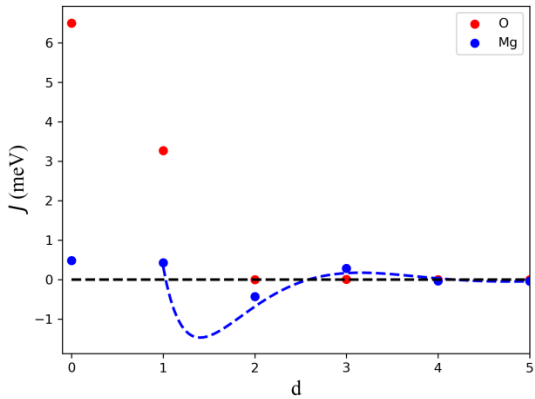

**Figure 3**. Exchange interaction J(d) for Mg- and O-edges as a function of distance d. Dots are calculated by linear response theorem and dashed line denotes the approximate RKKY solution (see eq. 11 in main text).

In Fig. 3, we show the calculated Heisenberg exchange parameters. Note, the on-site exchange $J(0)$ is defined as the summation of all exchange interactions: $J(0) = 2\sum_{d\neq 0} J(d)$, reflecting the averaged coupling effect, also equal to the energy difference between ferromagnetic and anti-ferromagnetic states: $J(0) = E_{FM} - E_{AFM}$. Both Mg- and O-edges have positive $J(0)$, indicating that these two edges are FM at least in ground state. For the O-edge, the FM exchange is largely contributed by the first nearest neighbor (1NN) interaction, while the second nearest neighbor (2NN) interaction is smaller by an order of magnitude and the sign also changes. The more distant interactions are negligible. These features indicate that the magnetism in O-edge has a very local nature, in consistent with the fact that the electronic states of O-edge are remarkably localized. Interestingly, the exchange interactions in Mg-edge show oscillatory features. There are dominating 1NN FM and comparable 2NN AFM coupling interactions, followed by decreasing magnitude and oscillatory signs for more distant interactions. This is similar to the RKKY interactions, which can be approximately expressed as[60]:

$$J(d) \propto \frac{\left|k_F^\uparrow + k_F^\downarrow\right| \sin\left[\left|k_F^\uparrow + k_F^\downarrow\right| d + \varphi\right]}{d^3} \qquad (11)$$

here $\varphi$ is the phase factor, $k_F^\uparrow$ and $k_F^\downarrow$ respectively denote the Fermi wavevectors for spin-up and down bands, i.e. the $Mg_2$ and $Mg_1$ bands marked in Fig. 2(c). This feature is not difficult to understand. As we discussed before, there are large portion of polar charges do not contribute to the local magnetic moment in Mg-edge but forming itinerant 1DEG. They spread between the local moments and transmit the exchange interactions, leading to the oscillatory nature. The significance is simply enormous since we have found that the Z-MgONRs support the existence of two kind of the magnetism with very distinct features.

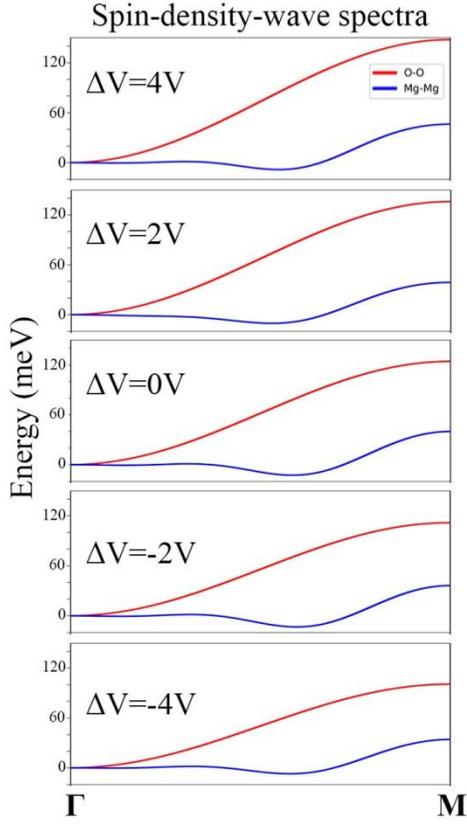

**Figure 4**. SDW spectra along high-symmetry path. Red and blue lines correspond to the O- and Mg-edge, respectively. ΔV denotes the applied biased voltage.

More detailed features of exchange interactions and the exotic spin-dynamics in Z-MgONRs can be revealed in terms of spin-density-wave (SDW) spectra. According to eq. 6, the dispersion law of SDW can be expressed through the reciprocal exchange parameter $J(k)$:

$$E(k) = \frac{1}{M}(J(0) - J(k)) \qquad (12)$$

where M is the magnitude of local magnetic moment, reciprocal $J(k)$ is related to its real-space version via lattice Fourier transformation: $J(k) = \sum_{d \neq 0} e^{ikda} J(d)$, here $a$ is the lattice constant of Z-MgONRs.

As can be seen in middle part of Fig. 4, there is no negative energy in the SDW spectra of O-edge, indicating that the FM order in O-edge is kinetically stable. As for Mg-edge, however, there are negative-energy troughs in the spectra. Apparently, these troughs are caused by the AFM exchange parameters J(2) and reveal that the long ranged AFM states are more energetically favored, while the FM order is just meta-stable in 0K. Once the temperature is beyond 0K with any infinitesimal magnitude, the magnetism in Mg-edge would spontaneously and irreversibly evolves to the AFM order as time elapsed.

### 3. Modulations of exchange interactions

Fig. 4 also shows a significant modulation of SDW spectra via biased voltage. For the O-edge, positively biased voltage makes the diversion relations steeper, indicating that the system becomes more likely to keep the FM order against thermal excitations. This result can be easily understood in terms of the DOS diagram. Positively biased voltage (corresponding to the **E** pointing upwards in Fig. 1(c)) pushes the $O_2$ peak closer to the Fermi level, leading to not only the reduction in hole-filling and magnetic moment as discussed former, but also the enhancement of DOS at Fermi level. According to Stoner criteria, higher DOS implies stronger exchange splitting, enhancing the 1NN FM exchange interactions in O-edge and making the SDW spectra steeper.

As for the Mg-edge, the electrical modulation of SDW spectra becomes bizarre. One can see from Fig. 2 that the Fermi wavevectors of bands of edge-Mg especially $k_F^\downarrow$ is clearly reduced (enhanced) by positively (negatively) biased voltage. In Fig. 3, we find that 1NN FM exchange interaction J(1) is lying at the monotonically decreasing region of sine part of eq. 11. Therefore, the reduction (enhancement) of Fermi wavevector $\left|k_F^\uparrow + k_F^\downarrow\right|$ caused by positively (negatively) biased voltage enhances (reduces) the sine part. The evolution of J(1) is thus determined by the competition between these two factors. This conclusion holds true for the 2NN AFM exchange interaction J(2), too. The complexities of exchange interactions give rise to the oscillatory characters in the modulation of SDW spectra. Under a positively biased voltage, the sine part always wins and the negative energy troughs are lifted up. When negatively biased voltage turns from zero to -4V, however, the sine part only wins in the first half, and the troughs become deep first, then become shallow.

The modulation of FM exchange interaction in O-edge indicates that the Curie temperature Tc may be also well tuned via **E**. In the framework of mean-field theory, Tc is proportional to the averaged exchange interaction: $(3/2)k_B T_C = J(0)$. According to eq. 12, J(0) is equal to the energy of SDW with wavevector locating at the edge of Brillouin zone, i.e. $(3/2)k_B T_C = E(k = \mathbf{M})$. The spectra in Fig. 4 show that positively biased voltage enhances this energy, we thus can crudely say that the positively biased voltage also enhance the Curie temperature of O-edge.

Strictly speaking, mean-field solution thoroughly neglects collective excitations which are the essential features for magnetic disorder at low temperature. Therefore, the Curie point would be drastically overestimated. For the sake to bring our study into a more accuracy level, we alternatively employ a series of Monte Carlo simulations to investigate the magnetic behaviors and electrical modulations in finite temperature. We assume strong magnetic anisotropy so that eq. 6 reduced to Ising model and we consider the magnetic coupling up to 3NN. When a FM system with infinite scale reaches Curie temperature, its susceptibility diverges. For our model with finite scale, the susceptibility will not diverge really, and we approximately regard the temperature which corresponds to the maximum of susceptibility as Tc, marking as black dot in Fig. 5(b). Along with the biased voltage increasing from -4V to 4V, Tc also increases from ~18K to ~23K, with about thirty percent enhancement. Fig. 5(a) shows the electrical modulation of magnetic moment near Tc. When the biased voltage positively increases, the ground-state magnetic moment will decrease as discussed before. On the other hand, the biased voltage also magnifies Tc thus suppresses thermal excitations. These two factors complete and finally lead to the magnetic moment increase first, following by a decreasing process, along with the increasing of positively biased voltage.

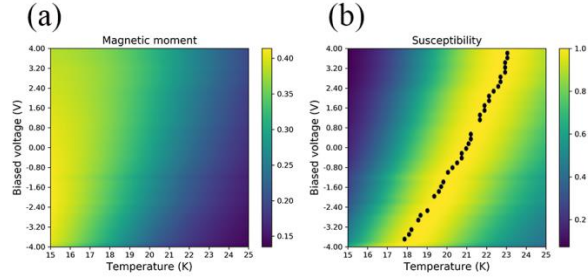

**Figure 5**. (a) magnetic moment (b) susceptibility patterns versus applied biased voltages and temperatures. The black dots in (b) denote the position of maximum susceptibility in each row.

## Conclusions

In summary, we have examined the magnetism and ME coupling mechanism in Z-MgONRs. It was demonstrated that the both Mg-edge and O-edge are magnetic and their **M** are dependent on polar charges and the spin-polarization of edge states. Since the polar charges are sensitive to the biased voltage, **M** thus can be efficiently modulated by **E**. In particular, O-edge is half-metallic and all polar charges are spin-polarized, **M** of O-edge will be enhanced by positively oriented **E**. Meanwhile, for Mg-edge, **E** will reduce the spin-polarization thus reduce **M**. Based on effective TB model and linear response theorem, we have revealed that the exchange interactions in O-edge are short range and FM, while the Mg-edge supports long range RKKY-like couplings. And these exchange interactions are also amenable to **E**, giving rise to the efficient electrical modulation of SDW spectra. Furthermore, a series of Monte Carlo simulations show that the Curie temperature of ferromagnetism in O-edge can also be well tuned by **E**. All these findings reveal the promising applications of Z-MgONRs in manifold spintronics, as well as give a comprehensive understanding of the magnetism and ME mechanism in Z-MgONRs.

## Conflicts of interest: no

## Acknowledgements

We acknowledge the inspiring discussions with X. B. Liu and the support of the National Natural Science Foundation of China (Grant Nos. 51472150, 51472145 and 51272133), Shandong Natural Science Foundation (Grant No. ZR2013EMM016), National 111 Project (B13029), and the Fundamental Research Funds of Shandong University (No. 2018GN037).